\documentclass[letter,printer]{aa}

\usepackage{caption}
\usepackage{graphicx}
\usepackage[position=top]{subfig}
\usepackage{amsmath}
\usepackage[utf8]{inputenc}
\usepackage{epsfig,amsmath,amssymb}
\usepackage[varg]{txfonts}
\usepackage{academicons}
\usepackage[breaklinks, colorlinks, citecolor=blue]{hyperref}

\def\ga{\,\hbox{\hbox{$ > $}\kern -0.8em \lower 1.0ex\hbox{$\sim$}}\,}
\def\la{\,\hbox{\hbox{$ < $}\kern -0.8em \lower 1.0ex\hbox{$\sim$}}\,}
\def\beq{\begin{equation}}
\def\eeq{\end{equation}}

\begin{document}

%
\title{A closer look at supernovae as seeds for galactic magnetization}
\authorrunning{Ntormousi et al.}
\titlerunning{Magnetization from supernovae?}
\author{Evangelia Ntormousi \orcid{0000-0002-4324-0034}\inst{1,2}, Fabio Del Sordo\orcid{0000-0001-9268-4849} \inst{3,4,5}, Matteo Cantiello\orcid{0000-0002-8171-8596} \inst{6} and Andrea Ferrara\orcid{0000-0002-9400-7312}\inst{1}} 
\date{Received -- / Accepted --}

\institute{
Scuola Normale Superiore,
Piazza dei Cavalieri, 7
56126 Pisa, Italy
\and
Institute of Astrophysics,
Foundation for Research and Technology (FORTH), 
Nikolaou Plastira 100, Vassilika Vouton
GR - 711 10, Heraklion, Crete, Greece
\and
Institute of Space Sciences (ICE-CSIC), Campus UAB, Carrer de Can Magrans s/n, 08193, Barcelona, Spain
\and
Institut d’Estudis Espacials de Catalunya (IEEC), 08034 Barcelona, Spain
\and
INAF, Osservatorio Astrofisico di Catania, via Santa Sofia, 78 Catania, Italy
\and
 Center for Computational Astrophysics, Flatiron Institute, 62 5th Ave, New York, NY 10010, USA
}

\abstract
{
Explaining the currently observed magnetic fields in galaxies requires relatively strong seeding in the early Universe. 
One of the current theories proposes that magnetic seeds of the order of $\mu$G were expelled by supernova (SN) explosions after primordial, nG or weaker fields were amplified in stellar interiors.
}
{In this work, we take a closer look at this theory and calculate the maximum magnetic energy that can be injected in the interstellar medium by a stellar cluster of mass $M_{cl}$ based on what is currently known about stellar magnetism.}
{We consider early-type stars and adopt either a Salpeter or a top-heavy IMF. For their magnetic fields, we adopt either a Gaussian or a bimodal distribution. The Gaussian model assumes that all massive stars are magnetized with $10^3 < \langle B_* \rangle < 10^4$~G, while the bimodal, consistent with observations of Milky Way stars, assumes only $5-10$ per cent of OB stars have $10^3 < \langle B_* \rangle < 10^4$~G, while the rest have $10 < \langle B_* \rangle < 10^2$~G.
We ignore the effect of magnetic diffusion and assume no losses of magnetic energy.}
%
{We find that the maximum magnetic energy that can be injected by a stellar population is between $10^{-10}-10^{-7}$ times the total SN energy. 
The highest end of these estimates is about five orders of magnitude lower than what is usually employed in cosmological simulations, where about $10^{-2}$ of the SN energy is injected as magnetic.
}
{
Pure advection of the 
stellar magnetic field by SN explosions is a good candidate for seeding a dynamo, but not enough to magnetize galaxies. Assuming SNe as main mechanism for galactic magnetization, the magnetic field cannot exceed an intensity of $10^{-7}$ G in the best-case scenario for a population of $10^{5}$ solar masses in a superbubble of 300 pc radius,
while more typical values are between $10^{-10}-10^{-9}$~G.
Therefore, other scenarios for galactic magnetization at high redshift need to be explored.}


\maketitle

\section{Introduction}
The origin of magnetic fields is one of the most compelling problems in astrophysics and cosmology. While some constraints of the primordial magnetic field strength exist, the mechanism that can generate them are very diverse \citep[e.g.,][]{gnedin2000,subramanian2019}. 
What we do know is that the magnetic fields currently observed in spiral galaxies are too strong to be primordial. Therefore, they were either seeded by a primordial field in the early Universe and then amplified through a dynamo, or directly generated by an astrophysical process in galaxies \citep[e.g.][]{dynreview2022}.

One theory for magnetizing early galaxies involves stars \citep{rees1987}. The idea is that a dynamo process in stellar interiors amplifies the tiny magnetic seeds coming from a Biermann battery or similar mechanism. When these stars explode as supernovae (SNe), the amplified fields are injected in the surrounding interstellar medium (ISM). 

A number of numerical experiments adopt this mechanism as a subgrid model in galactic dynamo \citep{Hanasz2009,Kulpa-Dybel_2011,Kulpa_dybel_2015,Butsky_2017} or cosmological structure formation simulations \citep{Beck_2013,Vazza2017,Katz2019,garaldi2021,MA2021}, injecting magnetic field together with thermal and kinetic
energy as feedback from stellar populations. 
However, the assumptions about the relative ratio of magnetic to thermal SN energy differ strongly between models. 
For instance, \citet{Kulpa-Dybel_2011,Kulpa_dybel_2015} assume that 10\% of the SNe introduce a field of the order of $10^{-5}~\mu$G
in dense regions of the ISM. A similar assumption on the incidence of SN magnetization is made by \citet{Hanasz2009}, although they do not specify the value of the magnetic field. \citet{Butsky_2017} seed each SN event with $10^{-7}$ of the SN energy as magnetic energy. Cosmological models typically assume that the contribution from SN is 
much
larger, and inject an energy $E_{\rm mag,inj}=\epsilon~E_{\rm SN}$ with $\epsilon$ typically ranging from 0.01 to 0.03. This implies field strengths of about $10^{-4}$~G on the typical scale of a remnant, which is about
5-10~pc. 
While these approaches provide a useful starting point for structure formation and cosmological simulations, they are not based on detailed models or observations of stellar magnetism.

In this Letter, we use our current understanding of stellar magnetism to provide a simple estimate of the maximum amount of magnetic energy that can be injected by a primordial stellar population. 
To our knowledge, this estimate has not been performed previously, and can be very useful for studies of cosmological magnetic field evolution.
We will show that the typical magnetic seeding by SN explosions used so far in cosmological simulations is much larger than what can be produced by a typical stellar population. 
Instead, the smaller values assumed by some galactic dynamo models are closer to our estimates.

\section{Stellar Magnetic Fields}

Equipartition between magnetic energy and gravitational binding energy sets the maximum allowed magnetic field $B_{\rm max}\propto M/R^2$ in a star. In early-type stars this maximum surface magnetic field is about $10^7$ G.  
On the other hand the largest observed fields at the surface of stars seem to be much smaller, with $B_{\rm max, obs}\sim \left(10^{-3}-10^{-4}\right) B_{\rm max}$ \citep{Braithwaite:2017}. This means that magnetic fields usually provide a very small perturbation to the stellar hydrostatic equilibrium. Also in early-type stars the maximum observed amplitude of surface magnetic fields is $10^3-10^4 $ G \citep{Petit:2019}. 

The origin and evolution of stellar magnetism is not fully understood. It is clear that dynamo action can amplify a seed magnetic field in stars, with the dynamo tapping into the energy reservoir provided by turbulent convection and/or differential rotation 
\citep{Bran2012}.
This is the case of the Sun and of low-mass stars in general.  On the other hand, the large scale, high amplitude fields observed at the surface of 5-10\% of early-type stars (``fossil fields'') are thought to be inherited during the star formation process \citep[e.g.][]{Donati:2009}. It is also possible that some of these fields are created via interactions with a companion star \citep{Ferrario:2009}. 
Finally, dynamo action might be at play in the interiors of early-type stars as well, both in their convective cores \citep{Augustson:2016}, near-surface convective zones \citep{Cantiello:2011}, and radiative regions \citep{Spruit:2002,Wheeler:2015,Fuller:2019}

As we outlined in the introduction, 
state-of-the-art cosmological models
and some galactic dynamo models
approach stellar magnetic seeding by introducing a magnetic field 
corresponding to roughly 1\% the SN energy
in feedback regions. 
The motivation behind this 
choice is based on estimates from galactic SN remnants. For instance, \citet{Beck_2013} estimate an average SN remnant of radius 5~pc and magnetic field strength $10^{-4}$~G, citing the review  of magnetism around local SN remnants by \citet{Reynolds2012}. This estimate corresponds to about $6\times10^{48}$~ergs in total magnetic energy.
However,
assuming that such strong magnetic fields at the SN stage come directly from the progenitor implies paradoxically high 
magnetization for the star. 
If the interstellar medium at the typical radius of a SN remnant $R_{\rm SN}$ contains $E_{\rm mag,inj}=0.01~E_{\rm SN}$, with $E_{\rm SN}=10^{51}$~ergs, then
the magnetic energy density of the progenitor would be $E_{\rm mag,inj}$ divided by the volume of the star:
\begin{equation}
    u_{\rm mag,*} = E_{\rm mag,inj}/V_*
\end{equation}
assuming no magnetic energy was lost or gained during the explosion.
For an O-type star with a radius $R_*=100~R_\odot$ and $E_{\rm mag,inj}=10^{49}$~ergs, $u_{\rm mag,*}\simeq 10^{10}$~ergs cm$^{-3}$. Since $u_{\rm mag} = B^2 / 8\pi$, the typical magnetic field of the progenitor would be roughly $B_*=10^5-10^6$~G. This number is at least one order of magnitude larger than the maximum magnetic field observed at the surface of early-type stars \citep[e.g.][]{Petit:2019}. 
This inconsistency is exacerbated assuming a smaller radius for the progenitor star, or dissipative losses during the evolution of the SN.

Clearly, the typical values for SN-injected magnetic fields used so far in cosmological simulations are unreasonably high, 
although they match the observed magnetic fields around local SN remnants, which are of the order of a few $\mu$G to a mG. However, there is currently no theory that involves magnetization of the SN ejecta to explain these values. Instead, current theories involve the amplification of the magnetic field \emph{surrounding} the remnant, through either compression and turbulence \citep{Inoue_2009}, or a cosmic-ray-driven dynamo \citep[e.g.,][]{xulazarian2017}.
In the following section we will attempt to make a physically meaningful estimate for magnetic seeding from SNe.

\section{The maximum SN-injected magnetic energy from a stellar population} 

We construct a simple model for the magnetization of a primordial cloud by SN ejecta. In this model, we assume that the stars
return all their magnetic field to the ISM. In doing so, we implicitly require that all the mass of the star is ejected in the ISM, neglecting the mass and magnetization of the resulting compact object. Obviously, this assumption is not true for most massive stars, but it helps our purpose of calculating an \emph{upper limit} to the magnetic field that can be injected by the population.
We also neglect dissipative losses or other forms of magnetic energy gain, so that the magnetic energy is conserved.

\begin{figure}
    \centering
    \includegraphics[width=0.9\linewidth]{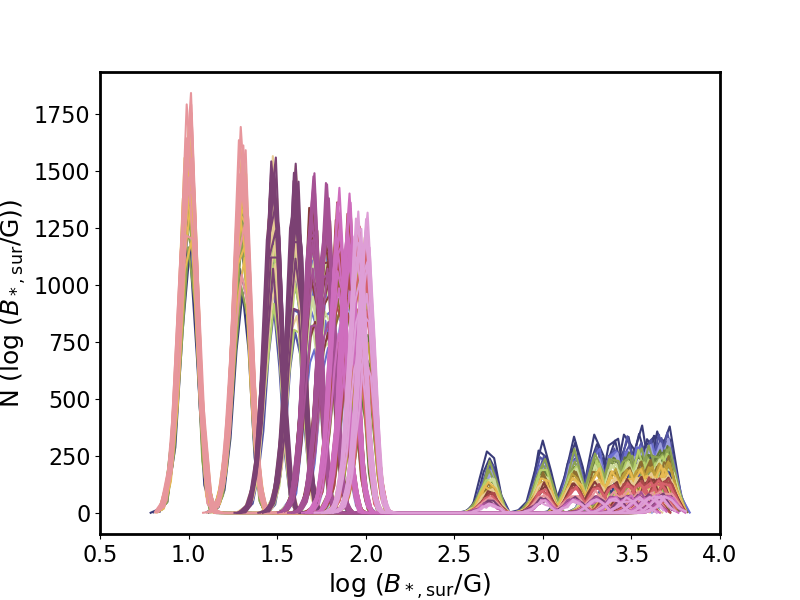} \\
    \includegraphics[width=0.9\linewidth]{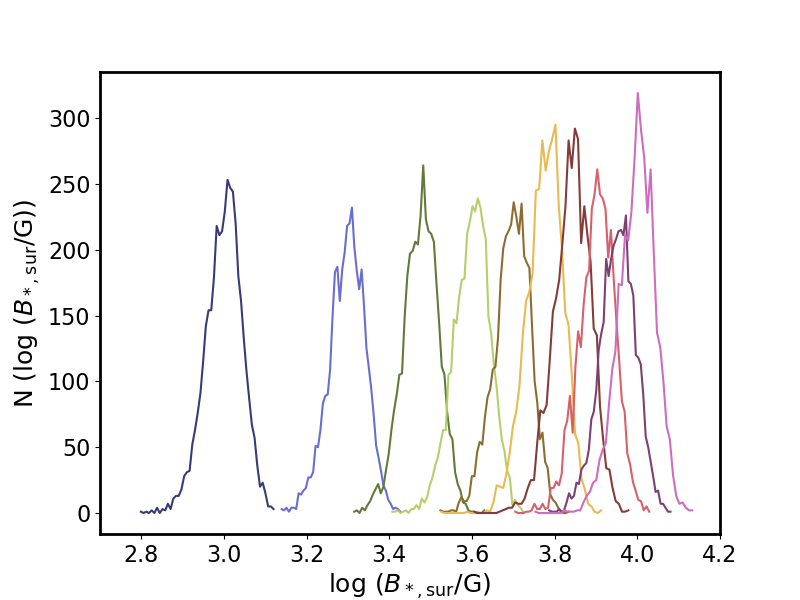}
    \caption{Range of 
    stellar surface
    stellar magnetic field distributions used in the two models. Top: bimodal, bottom: single Gaussian, both shown here for a cluster of $10^5$ $M_\odot$ and a Salpeter IMF.
    Each color corresponds to a set of parameters for each distribution: $(\mu_1,\mu_2,f_2)$ for the bimodal, $\mu$ for the Gaussian. The dispersion of each distribution with mean $\mu_i$ is set to $\sigma_i=0.1\mu_i$.}
    \label{fig:bofm}
\end{figure}

The model essentially consists of the following relation for the total magnetic energy generated by a cluster of mass $M_{\rm cl}$: 
\begin{equation}
    E_{\rm mag,cl} = \int_{M_{\rm SN}}^{M_{\rm max}(M_{\rm cl})} \xi(m) \left(\frac{B_*^2}{8\pi}\right)\left(\frac{4\pi R_*(m)^3}{3}\right) dm
\label{eq:magsn}
\end{equation}
where $\xi(m)$ is the assumed Initial Mass Function (IMF) of the primordial stellar population, $B_*$ the 
magnetic field of a star of mass $m$\footnote{$B_*$ does not depend on $m$ in this model, in agreement with the existing observations 
of massive stars
\citep{wade2015}}, $R_*(m)$ its radius, $M_{\rm SN}$ the minimum mass for SNe and $M_{\rm max}(M_{\rm cl})$ the maximum stellar mass of a cluster of mass $M_{cl}$.
We will estimate $E_{\rm mag,cl}$ for different masses of the stellar cluster.
All the quantities that enter Eq. (\ref{eq:magsn}) are only partially known, with $\xi(m)$ and $B_*$ the least understood so far. However, we can make order-of-magnitude estimates based on what is currently known.

\subsection{IMF of the star cluster}
The IMF of stars in primordial, low-metallicity environments is highly uncertain.
Numerical work \citep[e.g.,][]{Clarck_2008,Dopcke_2013,chon2021} has shown that for metallicities below $Z/Z_\odot\simeq10^{-5}$ fragmentation becomes inefficient and the stellar IMF favors high-mass stars, with an almost flat distribution at high masses. Above this metallicity limit, simulations recover a Salpeter-like power-law slope in high masses \citep{Salpteter_1955}. 
Therefore, here we 
will use two IMF models:
i) a Salpeter IMF: $\xi_{\rm S}(m) = \xi_0 m^{-2.35}$ 
(which for $m>0.5M_\odot$ coincides with the \citet{Kroupa_01} and \citet{Chabrier_03} models), 
and ii) a top-heavy IMF, with a high-mass slope of -1.5: $\xi_{\rm TH}(m) = \xi_0 m^{-1.5}$. 
As limits in the integration of the IMF we set $M_{\rm SN}=8~M_\odot$ and $M_{\rm max}(M_{\rm cl})=1.2\cdot M_{\rm cl}^{0.45}$ \citep{Larson2003}. 
Finally, for $R_*(m)$ we will use the simple relation for non-convective stars, $R_*(m)\propto m^{0.57}$.

\subsection{Stellar magnetic field distribution}
For the distribution of $B_*$ we use two different models,
starting from assumptions about the \emph{surface} magnetic field of the stars.
The first model is driven from local observations of early-type stars,
which show a very low incidence of detectable magnetization. Specifically, only about 5 to 10 percent of the observed stars host magnetic fields of
$10^3-10^4$ G
while the rest fall below about 100 G
\citep{Hubrig_2011,Alecian_2014,morel2014,wade_2014,Petit:2019}.
In fact, there are indications that OB stars present a magnetic dichotomy similar to that observed for Ap stars. Interestingly, the incidence of high magnetization does not seem to correlate with the stellar parameters \citep{wade2015}. 
Therefore, 
our first model approximates
the distribution of stellar surface magnetic fields, $B_{\rm *,sur}$, as a bimodal for all the stars contained in the integral of Eq. (\ref{eq:magsn}). The parameters of the distribution are the two means, $\mu_1$ and $\mu_2$, which we vary from 
$10~G<\mu_1<100~G$ and $500~G<\mu_2<5\times10^{3}~G$ 
respectively, and $f_2$. the incidence of highly magnetized stars, which takes values $0.05<f_2<0.2$. 

Recently, \citet{Farell2022} showed that the surface magnetic field distribution of AB stars is well fitted by a Gaussian with a mean at about 1kG. Although their study involved less massive stars, we explore a second model where all OB stars are strongly magnetized, in order to push our calculations towards an upper limit. In this case, $B_{\rm *,sur}$ is a single Gaussian with a peak that we vary between
$10^3<\mu<10^{4}$~G. 
The two panels of Fig. \ref{fig:bofm} illustrate the resulting range of $B_*$ distributions for the two models, for a stellar population of $10^5$ $M_\odot$. The different line colors correspond to different sets of parameters.

Going from the \emph{surface} to the \emph{global} field strength requires an assumption for the internal magnetic field distribution.
Here, we assume a dipole distribution of the magnetic field strength as a function of radius $r$:
\begin{equation}
    B*(r) = \int_{R_{\rm core}}^{R_*} 4\pi r^2B_{\rm *,sur}\left(\frac{R_*}{r}\right)^3 dr
    \label{eq:dipole}
\end{equation}
Since the dipole field diverges at the center, in Eq. (\ref{eq:dipole}) the integration starts from a core radius, $R_{\rm core}$. 
O- and B-type stars rotate rapidly and possess convective cores, which means that a dynamo could act in their interior \citep{Augustson:2016}. Although a core dynamo cannot be an explanation for the strong surface fields \citep[since there is no time for such a field to reach the surface -- ][]{Charbonneau2001}, a fossil field from a convective core could still be partly expelled with the SN.  
Equipartition fields in the convective cores of OB stars are expected to be in the range $10^5-10^6$~G \citep{Augustson:2016}.
With this in mind, we estimate $R_{\rm core}$ as the largest of two radii: i) the radius at which the magnetic field strength from Eq. (\ref{eq:dipole}) reaches the maximum stable field according to dynamo models \citep[e.g.,][]{Augustson:2016}, 
$B_*(R_{\rm core})=10^6$~G or ii) the mean core radius for massive stars, $R_{\rm core}=0.2R_{*}$\footnote{Varying the range of $R_{\rm core}=0.2-0.4R_{*}$ has a negligible effect on the results}. Inside the core the magnetic field is considered constant, either at the maximum value of $10^6$~G, or to the value reached at $R_{\rm core}$ from Eq. (\ref{eq:dipole}).

\subsection{Results}
\label{sec:results}
\begin{figure}
    \centering
    \includegraphics[width=\linewidth]{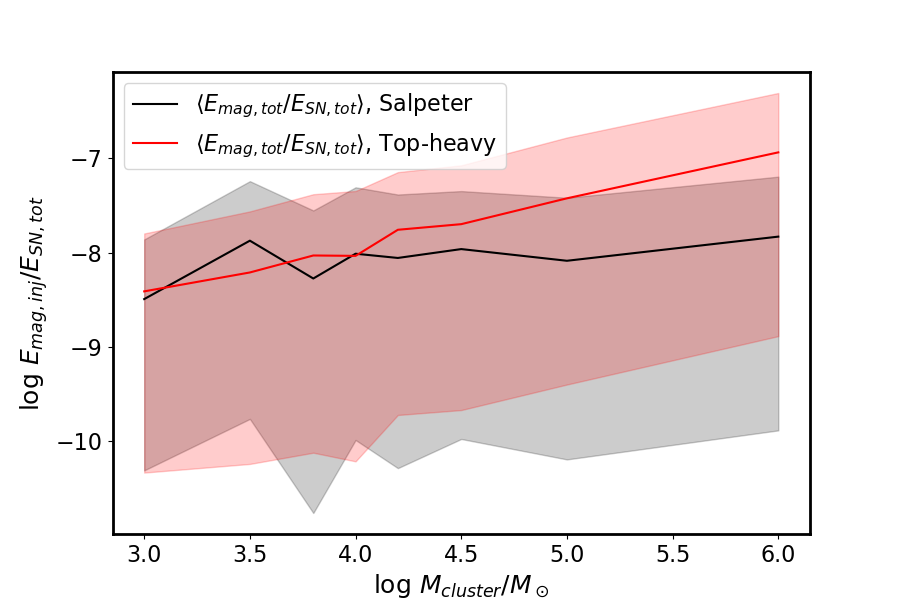}
    \includegraphics[width=\linewidth]{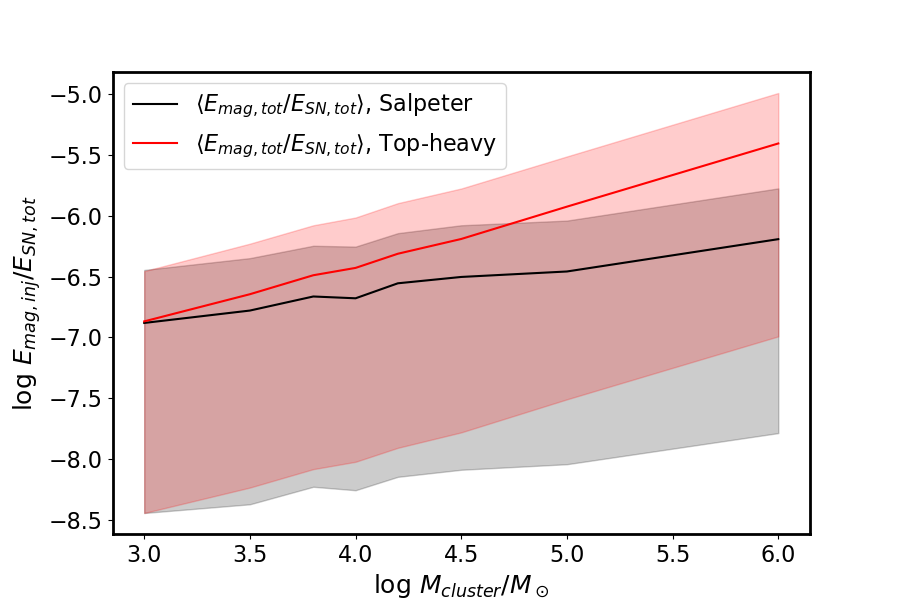}
    \caption{Magnetic energy injected by a stellar population of mass $M_{cluster}$ over SN energy of the same population. The latter is calculated as $10^{51}$~ergs times the total number of massive stars. The shaded area shows the range of values assuming different magnetic field distributions for the stars. The top panel assumes bimodal magnetic field distributions (top panel of Fig.\ref{fig:bofm}) and the bottom one Gaussian (bottom panel of Fig.\ref{fig:bofm}). The two lines in each panel show results for different IMFs.}
    \label{fig:mass_emag}
\end{figure}
\begin{figure}
    \centering
    \includegraphics[width=\linewidth]{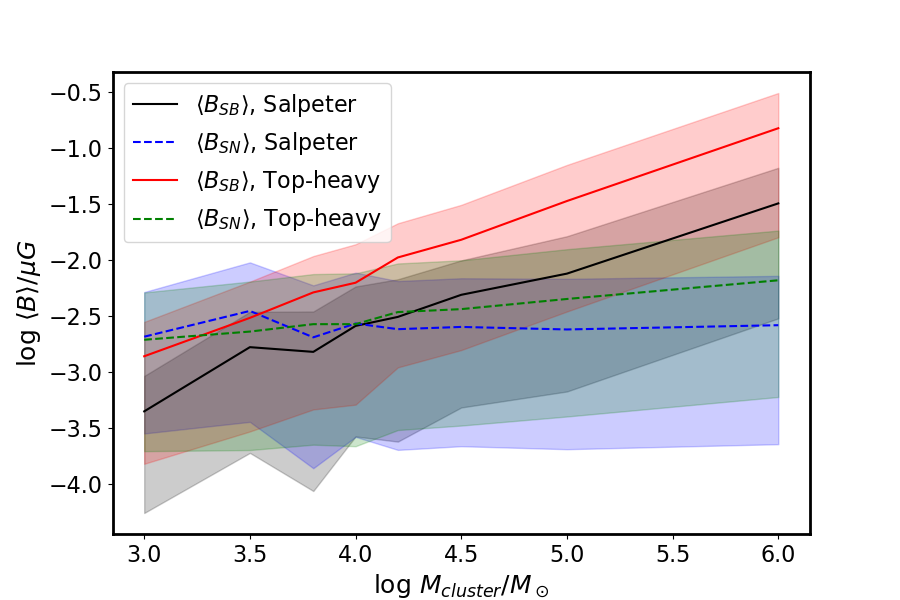}
    \includegraphics[width=\linewidth]{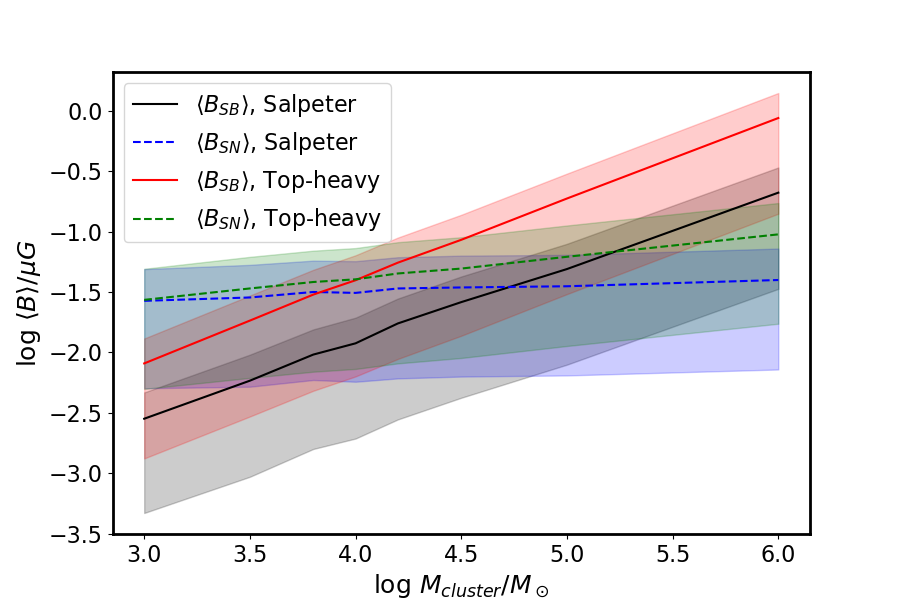}
    \caption{Total magnetic field in a single superbubble, $\langle B_{\rm SB}\rangle$, or mean magnetic field in SN remnants, $\langle B_{\rm SN}\rangle$, injected by a stellar population of mass $M_{\rm cluster}$. The top panel assumes bimodal magnetic field distributions (top panel of Fig.\ref{fig:bofm}) and the bottom one Gaussian (bottom panel of Fig.\ref{fig:bofm}). Results are shown for different IMFs, with the shaded areas indicating the range over different stellar magnetic field distributions.}
    \label{fig:mass_bfield}
\end{figure}

Figure \ref{fig:mass_emag} shows the result of this calculation in terms of the fraction of injected magnetic energy $E_{\rm mag,inj}$ over total SN energy $E_{\rm SN,tot}$. This quantity is equivalent to the factor $\epsilon$ employed in numerical simulations of magnetic field injection. Even at the highest limit of this range, assuming a top-heavy IMF and all the stars strongly magnetized (red line of the bottom panel), $\epsilon$ does not exceed $10^{-7}$. This strong upper limit value is still 
$10^5$ 
times lower than the number typically assumed in cosmological models.

Figure \ref{fig:mass_bfield} 
shows the same results in terms of the magnetic field strength. Here, we have made two calculations, corresponding to two approaches of distributing the SN ejecta. One assumes that all the stars deposit their magnetic energy in a single superbubble with a radius of 300~pc (so that their magnetic energies are added), and is shown as solid lines. The other, shown in dashed lines, assumes that each star explodes in isolation, forming a SN remnant of average radius of 30~pc (so that their magnetic energies are averaged). In reality, both scenarios are possible, depending on the clustering of the stars. Here, we see that an extremely massive cluster, with a top-heavy IMF and all its early-type, strongly magnetized stars depositing their magnetic field in a single superbubble of just a 300~pc radius, would yield a magnetic field of about $10^{-7}$~G. More typical values instead are between $10^{-10}-10^{-8}$~G, comparable to other primordial seeding mechanisms.

\section{Discussion}

We have shown that the magnetic fields used in cosmological simulations \citep[e.g.][]{Beck_2013,Vazza2017,Katz2019,garaldi2021,MA2021} under the SN-injection scenario are exceedingly high. 
However, in cosmic-ray-driven galactic dynamo simulations,
\citet{Hanasz2009} and \citet{Kulpa-Dybel_2011,Kulpa_dybel_2015} only introduce magnetization in 10\% of the SN events. This approach could be considered equivalent to our bimodal distribution of stellar magnetization,
with the employed values ($10^{-5}~\mu$G) lower than our average estimates.
\citet{Butsky_2017} use magnetic seeds with strengths close to those we predict in this Letter: they inject about $6\times10^{-7}$ of the total SN energy as magnetic energy. All the above works found these small values to be sufficient for driving a galactic dynamo.

The upper limits we calculate here are sensitive to the assumed magnetic field distribution in the stellar interior and to the maximum allowed surface magnetic field strengths. Assuming a constant magnetic field, $B_*=B_{\rm sur}$, in the entire volume of the star instead of a truncated dipole lowers our estimates for the magnetic energy injection by roughly an order of magnitude. Allowing for extreme surface magnetic fields of the order of a few times $10^5$~G increases the maximum magnetic yield by roughly a factor of 200. 

We note that our model does not take into account any dissipative process. Moreover, we do not consider the magnetization of the compact objects left behind by the SNe: they could retain a significant amount of the magnetic energy assumed to be injected into the ISM.
%


Finally, here we have examined the effect of the short-lived, explosive, early-type stars on magnetic field seeding. However, stellar magnetic field injection is also possible through a dynamo-active accretion disk around a protostar.
These dynamo-generated fields are strong enough to launch a wind, which in turn could magnetize the ISM \citep[e.g.,][]{Bran2000,vrekowsky2003,Dyda2018}.
Such disks are long-lived for low-mass stars, which can keep seeding the ISM with low levels of magnetization on small scales.
This scenario is worth investigating with dedicated numerical models. 

\section{Conclusions}
\label{sec:concl}

We conclude that stellar magnetic fields purely advected through SN explosions can tap only  $10^{-10}-10^{-7}$ of the SN explosion energy.  
The highest limit of these estimates is five orders of magnitude lower than what is currently used in cosmological simulations, even without accounting for diffusion or a residual magnetization of the compact object.
Therefore, SN-injected magnetic fields can only provide the seed for additional processes, such as a turbulent dynamo, that can amplify them to the near-equipartition values measured in local galaxies. Since early galaxies do host strong turbulence, this scenario can be probed with high-resolution simulations of these systems.
The obvious next step for this investigation are resolved numerical simulations of SN-injected fields that can account for magnetic diffusion and model possible dynamo action.


\begin{acknowledgements}
We acknowledge the Nordita program on ``Magnetic field
evolution in low density or strongly stratified plasmas'' in May 2022, for hosting fruitful discussions that nourished the ideas behind this work. We are particularly grateful to Axel Brandenburg for useful discussions. 
EN and AF acknowledge funding
from the ERC Grant “Interstellar” (Grant agreement 740120)
EN also acknowledges funding from the HFRI's Second call for supporting post-doctoral researchers (Project number 224). 
FDS acknowledges support from a Marie Curie Action of the European Union (Grant agreement 101030103) and ``Mar\'ia de Maeztu'' award to the Institut de Ciències de l'Espai (CEX2020-001058-M).
The code used to produce the results in this letter can be found at: https://github.com/entorm/snseeding

\end{acknowledgements}

\bibliographystyle{apj}
\bibliography{supernova}
\label{lastpage}

\clearpage

\end{document}